\documentclass[%
 reprint,
nofootinbib,
 amsmath,amssymb,
 aps,
 prd
]{revtex4-1}

\usepackage{multirow}
\usepackage{amssymb}
\usepackage{amsmath,enumerate,mathtools}
\usepackage{amsthm}
\usepackage{wasysym}
\usepackage{enumitem}
\usepackage{tensor}
\usepackage{xcolor}
\usepackage{mathrsfs}
\usepackage{graphicx}
\usepackage{dcolumn}
\usepackage{bm}

\newtheorem{theorem}{Theorem}

\newtheorem{propos}[theorem]{Proposition}

\newtheorem*{prop*}{}

\theoremstyle{definition}

\DeclareMathOperator{\Tr}{Tr}

\newcommand*\DE{\ensuremath{\mathscr{D}}}

\newcommand*\de{\ensuremath{\textnormal{d}}}

\begin{document}

\title{Einstein-Yang-Mills fields immune to quantum corrections}

\author{Martin Kuchynka}
\email[]{kuchynkm@gmail.com}
\affiliation{Institute of Mathematics of the Czech Academy of Sciences, \v Zitn\' a 25, 115 67 Prague 1, Czech Republic\\
 Institute of Theoretical Physics, Faculty of Mathematics and Physics, Charles University in Prague, \\V Hole\v{s}ovi\v{c}k\'ach 2, 180 00 Prague 8, Czech Republic}

\begin{abstract}

Just recently, the class of all Einstein-Maxwell fields solving simultaneously also any higher-order modification of the Eintein-Maxwell theory has been completely identified. 
In the present work, we argue that, in view of our recent results on nonabelian gauge fields, analogous identification can be in fact achieved for Einstein-Yang-Mills fields associated with any compact and semisimple gauge group.  
In particular, any such solution consists of a $VSI$ (vanishing scalar invariant) spacetime metric and a $VSI$ gauge field, both subject to a simple tensorial condition. 
Based on that, we are able to provide explicit form of all such solutions and interpret them as gravitational and Yang-Mills plane-fronted waves propagating in flat spacetime along the common recurrent wave vector. 
The results have consequences also for theories with a richer field content, which is further illustrated on 10D heterotic supergravity.  

\end{abstract}


\maketitle

\section{Introduction \& Summary}
\label{intro}

Not long ago, certain $pp$-waves were the only metrics known to be resistant to any higher-curvature modifications of the Einstein field equation \cite{guven,klimcik,horowitz}, serving thus as "universal" exact solutions to any higher-order gravity theory. 
That was until Gibbons et al. \cite{metricsvanishing} noticed that in the absence of matter fields, these $pp$-waves share the universal property with a broader class of Ricci-flat plane-fronted waves with a \textit{recurrent} wave vector $\ell$ (i.e. such that $\nabla_\mu \ell_\nu \propto \ell_\mu \ell_\nu$) and soon after, vacuum metrics with vanishing higher-order gravitational corrections were found even beyond the recurrent class \cite{typeIIINuniversal}. 
Since then, electromagnetic fields exhibiting analogous universal property\footnote{Form fields immune to higher-order modifications of their equations of motion have been of interest mainly in effective theories of superstrings  \cite{guven,horowitz}, where the first examples of universal forms come from.} have also become a subject of a more systematical study \cite{VSIelmag,quantumelmag,universalMaxwell}. 
Aforementioned works on universal spacetimes and electromagnetic fields subsequently enabled a complete characterization of Einstein-Maxwell fields immune to any higher-order corrections \cite{EMuniversal} and initiated effort to investigate situation in the territory of nonabelian gauge fields.\footnote{Besides universality of Yang-Mills plane waves coupled to appropriate plane-wave metrics \cite{guven}, not much was known for a general nonabelian case.}

First steps in a more systematical analysis of universality in the context of nonabelian gauge theories were made in \cite{YMuniversal} by investigation of gauge fields with vanishing scalar curvature invariants ($VSI$ gauge fields, for short) and Yang-Mills test fields immune to any higher-curvature modification of the Yang-Mills equation. 

In the present contribution, we wish to complete the discussion of \cite{YMuniversal} by taking into account the spacetime backreaction and draw some general conclusions based on results of the previous works.

Section \ref{results} thus deals with universal solutions of the full Einstein-Yang-Mills (EYM) theory. These are, in particular, resistant to all higher-curvature corrections emanating from rather general action $S$ specified in Section \ref{actionsection}. 
Such corrections to the EYM theory appear most notably in the context of the string theory, where, apart from the gravitational corrections, nonabelian generalization of the Born-Infeld Lagrangian also emerges in the string effective action \cite{BITseytlin}. But they appear in a variety of other contexts as well, such as Yang-Mills hierarchies \cite{YMlovelock2} proposed as a natural higher-dimensional analogues of the Yang-Mills theory or various non-minimal extensions of the EYM theories \cite{nonminYM2,nonminYM1}.  

In a nutshell, the main result (Theorem \ref{main}) can be stated as follows:
\textit{An Einstein-Yang-Mills field $(g_{\mu \nu},A_{\mu})$ with a nonvanishing curvature $F_{\mu \nu}$ is immune to any higher-order corrections of $S$ if and only if both fields are $VSI$ and satisfy} 
\begin{equation}\label{conditions}
R_{\mu \rho \sigma \lambda}R\indices{_{\nu}^{\rho \sigma \lambda}} = 0, \qquad \Tr \DE_\rho F_{\mu \sigma} \DE^\rho F\indices{_\nu^{\sigma}} = 0. 
\end{equation}
More insight into the nature of these solutions can be gained by employing the characterization of $VSI$ metrics and $VSI$ gauge fields\footnote{Full characterization of $VSI$ gauge fields was obtained for the case of a compact and semisimple gauge group. For a general finite-dimensional gauge group, the corresponding conditions were proved to be sufficient, although they may not be necessary.} provided in \cite{VSIinHD,HDVSI} and \cite{YMuniversal}, respectively.  In particular, $g$ belongs to a subclass of Weyl type III and Ricci type\footnote{In four dimensions, these correspond to Petrov type III and Petrov-Plebański type O. For higher-dimensional algebraic classification of the Weyl and the Ricci tensors, see e.g. review paper \cite{review}.} 
N Kundt metrics and $A$ possesses a null field strength $F$.  Both fields are then aligned with a recurrent null  vector $\ell$, being thus a geodetic null vector with zero twist, shear and divergence. All $(g,A)$ immune to higher-order corrections thus consist of plane-fronted gravitational and Yang-Mills waves propagating in flat spacetime along the common recurrent wave vector. 
Consequently, coming back to issue of metrics with vanishing gravitational corrections, solutions presented in \cite{EMuniversal} and here demonstrate that some non-$pp$-wave metrics enjoy this property even in the presence of suitable matter fields. 
However, in contrast with the vacuum case, once a massless scalar field, $p$-form or gauge field is present, recurrence of the wave vector turns out to be mandatory.\footnote{The corresponding quadratic condition of type \eqref{conditions} fails to be compatible with null matter fields aligned with a nonrecurrent null vector. See Remark 3.2 of \cite{EMuniversal} and the discussion following Proposition \ref{necYM} in section \ref{results}.}

After employing adapted coordinates and restating conditions \eqref{conditions} in terms of the field's coordinate components, we are able to provide explicit form of the fields (section \ref{localform}). The line element takes the form \eqref{VSImetric} while the gauge field can always be cast in the form \eqref{localgauge} by a suitable gauge transformation. The fields are then subject to equations \eqref{Weylcond} - \eqref{-2ricci}. 
We observe that these solutions generalize nonabelian plane waves coupled to gravitational $pp$-waves\footnote{The solutions of \cite{guvenEYM} are four-dimensional, but can be straightforwardly extended to arbitrary dimension.} discovered in \cite{guvenEYM}. Those coincide with Weyl type N subclass of solutions with a gauge field \eqref{localgauge} having functions $f(u)$ and $f_i(u)$ bounded, in which case $\ell$ is necessarily covariantly constant, and thus a null Killing vector.

Lastly, some consequences beyond EYM theories are discussed. We observe that the approach of \cite{EMuniversal} and of this paper can be straightforwardly carried over to more general second-order theories of coupled gravity, gauge fields and forms. This is demonstrated in section \ref{consequences} on a particular example - the bosonic part of 10D heterotic supergravity \cite{chapline, guven}. In analogy with the pure EYM case, we conclude that supergravity solutions consisting of $VSI$ fields and subject to quadratic conditions analogous to \eqref{conditions} are again immune to any polynomial higher-order corrections to EOM. 
These solutions are closely related to G{\"u}ven's plane waves \cite{guven} with vanishing heterotic string $\alpha^\prime$ corrections, as they consist of the same dilaton, axion and Yang-Mills field but admit a broader class of metrics possessing a recurrent null vector $\ell$. 
In particular, also non-$pp$-wave backgrounds are included, albeit these don't preserve supersymmetry \cite{farrill} as $\ell$ fails to be a Killing vector in this case \cite{VSIsugra}. The subset of solutions having half of the supersymmetries unbroken then corresponds to solutions with Weyl type N metrics. 
The issue of $\alpha^\prime$ corrections for the extended class of solutions is briefly addressed at the end of section \ref{consequences}.

\section{Higher corrections to Einstein-Yang-Mills}\label{actionsection}
Let $g$ be a Lorentzian metric on a spacetime manifold of dimension $D >2$ and let $A$ denote a gauge field associated with a compact and semisimple gauge group. 
We consider a rather broad class of theories describing their (possibly nonminimal) interaction and for which the field equations take the form of the EYM equations corrected by higher-order terms. More precisely, any admissible action for $g$ and $A$ takes the form
\begin{equation}\label{action}
S =  \int \de^D x \frac{\sqrt{-g}}{16\pi} \bigg\{  R-2\Lambda  - \frac{\kappa_0}{2} \Tr F_{\mu\nu}F^{\mu \nu} + \mathcal{L}_{HC} \bigg\},
\end{equation}
where $\kappa_0$ is a coupling constant, $F \equiv \de A - i[A,A]$ and $\mathcal{L}_{HC}$ is the part of the Lagrangian representing higher-order corrections to the EYM theory. 
$\mathcal{L}_{HC}$ is assumed to be an analytic function of scalar polynomial invariants constructed from the Riemann tensor $\mathcal{R}$, its covariant derivatives of arbitrary order, the field strength $F$ and its gauge covariant derivatives of arbitrary order. Moreover, its Taylor series consists strictly of monomials of order\footnote{Following terminology of \cite{fkwcpaper}, by order of a tensor, we mean the total number of derivatives of the metric and the gauge potential involved, cf. also \cite{EMuniversal}.} greater than two, ensuring thus higher-order nature of the corrections to the second-order EYM Lagrangian.

For the sake of further discussion, one may decompose $\mathcal{L}_{HC}$ into the individual corrections 
$\mathcal{L}_{g}, \mathcal{L}_{A} $ and $\mathcal{L}_{int}$ to gravity, gauge theory and their possible nonminimal interaction\footnote{$\mathcal{L}_{int}$ consists of all the terms involving mixed invariants and hence cannot be put neither in $\mathcal{L}_{g}$ nor $\mathcal{L}_{A}$.}, respectively:
\begin{align}
\begin{split}
\mathcal{L}_{HC} \equiv\  &\mathcal{L}_{g}(\mathcal{R},\nabla \mathcal{R}, \dots) +  \mathcal{L}_{A}(F,\DE F, \dots)\\
 &+ \mathcal{L}_{int}(\mathcal{R},F,\nabla \mathcal{R},\DE F, \dots),
\end{split}
\end{align}
giving rise to actions $S_g$, $S_A$ and $S_{int}$ associated with the corresponding type of corrections in $\mathcal{L}_{HC}$. 
Here, $\nabla$ and $\DE = \nabla - i[A,\cdot]$ denote the metric compatible, torsion-free covariant derivative and gauge covariant derivative, respectively.

\section{Solutions with vanishing corrections}\label{results}
Now that theoretical setting for generalizations of EYM theories is established, we can discuss conditions under which the individual types of corrections associated with $\mathcal{L}_{HC}$ vanish and thus complete discussion started in \cite{YMuniversal}, where only corrections $\sim \delta S_A / \delta A^\mu$ to the Yang-Mills equation for test gauge fields were considered. 

In full theory \eqref{action}, a spacetime reacts to the presence of a gauge field via the stress-energy tensor, within which the standard Yang-Mills stress-energy tensor receives higher-order corrections emanating from the metric variation $ \delta S_A/\delta g^{\mu \nu}$. 
First, by considering suitable forms of Lagrangian $\mathcal{L}_A$, an observation analogous to the one in \cite{EMuniversal} can be immediately made:
\begin{propos}\label{necYM}
Any gauge field $A_{\mu}$, for which all higher-order corrections $\delta S_A/\delta g^{\mu \nu}$ of $S$ to the Yang-Mills stress-energy tensor vanish, is necessarily $VSI$.
\end{propos} 
In particular, $F$ is a null field strength living in a degenerate Kundt spacetime $g$ and propagates along a geodetic null vector $\ell$ with zero shear, twist and divergence \cite{YMuniversal}. 
By the same token, the requirement of vanishing gravitational corrections $\delta S_g/\delta g^{\mu \nu}$ constricts $g$ to be $VSI$ \cite{EMuniversal}. 
This means that the degenerate Kundt spacetime $g$ (which is in general of Weyl and Ricci type II \cite{hervikKundt}) needs to be further restricted to Weyl and Ricci type III in order not to suffer any gravitational corrections. At this moment, it is worth noting that once both $g$ and $A$ are $VSI$, all possible corrections $\delta S_A/ \delta A^\mu$ to the Yang-Mills equation already vanish due to Theorem 4.5 of \cite{YMuniversal}.

However, neither $g$ nor $A$ being $VSI$ is sufficient for the other types of corrections to vanish as counterexamples can be easily constructed. 
In fact, the requirement of vanishing variations of $R_{\mu \nu}R^{\mu \nu}$ and the Gauss-Bonnet term yields further nontrivial restrictions on $VSI$ Einstein-Yang-Mills solutions - these are precisely the conditions \eqref{conditions}. 
To be more precise, the first condition of \eqref{conditions} is nontrivial only in $D>4$ as it originates from the metric variation of the Gauss-Bonnet term, which vanishes identically in $D=4$.
The second condition of \eqref{conditions}, which is related via the Einstein equation to the variation of $R_{\mu \nu} R^{\mu \nu}$, turns out to be very restrictive on geometry of the solutions for two reasons. 
Firstly, in conjunction with the $VSI$ condition for $A$, it causes the curvature $F$ to be constant over the gauge fields's wave fronts (as will be seen in section \ref{localform}). 
Secondly, it requires the wave vector $\ell$ to be recurrent - only then the condition can be satisfied by a nonvanishing null Yang-Mills curvature, cf. also remark 3.2 of \cite{EMuniversal}.

We have thus seen how the $VSI$ property and the quadratic conditions \eqref{conditions} naturally arise solely from the requirement of vanishing corrections to the Einstein equation.
Interestingly enough, once these necessary conditions are met, any other possible corrections associated with $S_g$, $S_A$ or $S_{int}$ already vanish, arriving thus at the main result

\begin{theorem}\label{main}
Let $(g_{\mu \nu},A_\mu)$ be a solution of the Einstein-Yang-Mills theory with a nonvanishing field strength $F_{\mu \nu}$. Then, all higher-order corrections of $S$ to the Einstein-Yang-Mills equations vanish for $(g_{\mu \nu},A_{\mu})$ if and only if
both fields are $VSI$ and satisfy conditions \eqref{conditions}.
\end{theorem}
Indeed, employing identities (2.10), (2.11) of \cite{YMuniversal}, Lemmas B.6, B.7, D.3 and D.8 of \cite{EMuniversal} can be straightforwardly extended to Yang-Mills curvature $F$. The proof of the sufficient part of the theorem can be then carried along the lines of proofs of Theorems 3.1 and 3.4. We thus refer the reader to \cite{EMuniversal} for any technical details. 

Let us conclude the section with a few remarks. Firstly, note that Theorem \ref{main} holds true even for minimally coupled theories (with $\mathcal{L}_{int} = 0$) as corrections associated with $S_{int}$ were not needed to conclude $(g,A)$ have to be $VSI$ and subject to the conditions \eqref{conditions}. 
Moreover, solutions immune to higher-order corrections exist only for theories \eqref{action} with $\Lambda = 0$ (recall that both fields are $VSI$). 
Let us also note that, while $A$ is a special case of universal Yang-Mills fields of \cite{YMuniversal}, $g$ belongs to the class of the so-called \textit{almost universal} metrics introduces in \cite{Gorilla}. 
In particular, $g$ is an example of $TNS$ metrics (as defined in \cite{Gorilla}) with $\nabla_\rho \nabla^\rho R_{\mu \nu} = 0$ and can be related 
to appropriate recurrent Ricci-flat universal metric $g_U$ of \cite{typeIIINuniversal} via the generalized Kerr-Schild transformation $g=g_U + 2 \mathcal{H}\ell \otimes \ell$ with a suitable function $\mathcal{H}$ \cite{EMuniversal}. In the Weyl type N case, $g$ can be, in fact, always related by such a transformation to the Minkowski metric, being thus a Kerr-Schild spacetime (see remark 4.2 of \cite{EMuniversal}). Significance of such Kerr-Schild Kundt metrics in the context of generic gravity theories has been previously recognized in \cite{gurses2013}.

\subsection{Explicit form of the solutions in adapted coordinates}\label{localform}
From Section IV of \cite{EMuniversal} and Section 3.1 of \cite{YMuniversal}, we conclude that, in a suitable gauge,   
local form of a general Einstein-Yang-Mills solution of Theorem \ref{main} in adapted Kundt coordinates $(r,u,x^i)$ with $r$ being the affine parameter of the corresponding Kundt vector $\ell = \partial_r$ can be expressed as
\begin{equation}\label{VSImetric}
\de s^2 = 2 \big( r H^{(1)} + H^{(0)} \big)  \de u^2 + 2 \de u\left( \de r + W_i \de x^i \right) + \left(\de x^i \right)^2 ,
\end{equation}
\begin{equation}\label{localgauge}
A_\mu \de x^\mu =\left[ f_i (u) x^i + f(u) \right] \de u,
\end{equation}
 with $i,j = 2, \dots , D-1$ and $H^{(0)}, H^{(1)}, W_\alpha$ being functions of $u$ and $x^j$ only.
These are then subject to the following constraints
\begin{equation}\label{Weylcond}
W_{[i,j]k}W^{[i,j]k} = 
2 W\indices{_{[k,m]}^{m}}W\indices{^{[k,n]}_{n}},
\end{equation}
\begin{equation}\label{-1ricci}
H\indices{^{(1)}_{,j}} = W\indices{_{[j,k]}^{k}},
\end{equation}
\begin{align}\label{-2ricci}
\begin{split}
 \Delta H^{(0)} =\  &2 H\indices{^{(1)}_{,k}}W^{k} + H^{(1)}W\indices{^{k}_{,k}} + {W_{m,u}} \indices{^{m}}  \\  
 &+ W_{[m, n]} W^{[m , n]} - \kappa_0 \Tr f_{i} f^{i},
 \end{split}
\end{align}
where $\Delta$ is the Laplace operator in the flat transverse space. 
Indeed, a $VSI$ metric satisfying the corresponding Einstein equation and subject to $R_{\mu \rho \sigma \lambda}R\indices{_{\nu}^{\rho \sigma \lambda}} = 0$ takes the form 
\eqref{VSImetric} subject to \eqref{Weylcond}, \eqref{-1ricci} and \eqref{-2ricci} \cite{EMuniversal}, while the gauge freedom enables one to set $A_r,A_{i},A_{u,r}=0$ in the case of $VSI$ gauge field $A$ \cite{YMuniversal}. The condition $\Tr \DE_\rho F_{\mu \sigma} \DE^\rho F\indices{_\nu^{\sigma}} = 0$ then reduces to $F_{ui,j}=0$ (and consequently  $A_{u,i j} = 0$), arriving thus at \eqref{localgauge} with $f$ and $f_i$ being arbitrary Lie algebra-valued functions of $u$.

The above solutions are in general of Weyl type III and the vector $\ell$ is recurrent. In four dimensions, type III solutions with a null gauge field were first considered in \cite{typeIIIEYM}, containing the above solutions as a special case. 
The $pp$-wave ($\nabla \ell = 0$) subclass is characterized by $H^{(1)} = 0$. In $D=4$, this subclass is necessarily of type N and coincides (for $f$ and $f_i$ bounded) with solutions discovered in \cite{guvenEYM}. In higher-dimensions, the metrics are in general of type III, while type N subset is recovered by $H^{(1)}=0=W_{i}$ (it is thus a proper subset of $pp$-wave solutions in $D>4$). Equations \eqref{Weylcond}, \eqref{-1ricci} are then automatically satisfied, while \eqref{-2ricci} reduces to Poisson equation
\begin{equation}\label{Poisson}
\Delta H^{(0)} = - \kappa_0 \Tr f_i f^i.
\end{equation} 
Type N solutions are thus a higher-dimensional generalization of the ones in \cite{guvenEYM}.
Finally, Weyl type O (conformally flat) solutions correspond to the Weyl type N subcase with 
\begin{equation}\label{WeylO}
H^{(0)} = - \frac{\kappa_0 \Tr f_i f^i}{2(D-2)}\sum_j (x^j)^2.
\end{equation} 
See the discussion of Section IV in \cite{EMuniversal} for more details.

\section{Consequences beyond Einstein-Yang-Mills}\label{consequences}

It is not hard to see that approach of \cite{EMuniversal} can be just as well applied to richer theories involving a spacetime metric, gauge fields, scalars and $p$-forms. 
For the sake of definiteness, let us illustrate it on a particular example of a theory encompassing all these fields - the bosonic part of $D=10$ heterotic supergravity \cite{chapline, guven}
\begin{align} \label{sugraaction}
\begin{split}
S =\ & \int \de^{10} x  \frac{\sqrt{-g}}{2\kappa^2} \bigg( 
R  - \frac{\kappa^2}{2 \phi} F\indices{^a_{\mu \nu}} F^{a \mu \nu} \\ 
& -\frac{2}{ \phi^2}\partial_\mu \phi \partial^\mu \phi  + \frac{3\kappa^4}{\phi^2} H_{\mu \nu \rho} H^{\mu \nu \rho}
\bigg).
\end{split}
\end{align}
Due to G{\"u}ven's work \cite{guven}, exact plane wave solutions of \eqref{sugraaction} are known to be immune to any possible heterotic string $\alpha^\prime$ corrections and, in particular, to all polynomial higher-order corrections to the field equations constructed from the "field strengths" $\mathcal{R}, F, \nabla \phi, H$ and their derivatives of arbitrary order. 
These plane waves consist of: 
\begin{enumerate}[label=(\arabic*)]
\item[(g)] 10D metric \eqref{VSImetric} with $H^{(1)},W_i = 0$ and $H^{(0)}$ quadratic in $x^i$;
\item[(A)] gauge field \eqref{localgauge} associated with $SO(32)$ or $E_8 \times E_8$ gauge group;
\item[($\phi$)] dilaton $\phi = \phi(u)$;
\item[(B)] 2-form $B=b_i(u,x^j) \de x^i \wedge \de u$ with $b_i$ linear in $x^j$;  
\end{enumerate}
See equations (10)-(14) and  (17) - (20) of \cite{guven}.

\subsection{Extending the class of solutions immune to higher-order corrections} \label{extending}
Based on the discussion in \cite{EMuniversal} and above, the conclusion of vanishing higher-order corrections to the field equations can be extended to any Ricci-nonflat\footnote{In fact, it is sufficient to assume that at least one of the fields $\textnormal{Ric},F,\nabla \phi, H$ does not vanish.} $VSI$ solution $(g,A,\phi,B)$ of \eqref{sugraaction} such that the following tensors vanish
\begin{equation}\label{quadratictensors}
\begin{gathered}
R\indices{_{\mu \rho \sigma \lambda}}R \indices{_\nu^{\rho \sigma \lambda}}, \qquad 
\Tr \DE_\rho F\indices{_{\mu \sigma}} \DE^\rho F\indices{_\nu^{ \sigma}}, \\
\nabla_{\rho} \nabla_\mu \phi \nabla^\rho \nabla_\nu \phi, \qquad
\nabla_\rho H\indices{_{\mu \sigma \lambda}} \nabla^\rho H \indices{_\nu^{\sigma \lambda}}.
\end{gathered}
\end{equation}

Solutions just described consist of G{\"u}ven's plane wave fields $A$, $\phi$ and $B$ coupled with a more general metric $g$ given by \eqref{VSImetric} and satisfying appropriate modification of the Einstein equation \eqref{-2ricci} (the equation \eqref{-2ricci2} of the next section). 
Whether all heterotic $\alpha^\prime$ corrections remain zero even for this broad class of solutions remains an open problem.\footnote{The situation is completely different once we restrict to $pp$-wave metrics, in which case a variety of well-known results is available, see e.g. \cite{hortsey, TseytlinDuality} and references therein.}
The main obstacle in applying the approach of \cite{EMuniversal} arises from the fact that the string effective action (as well as the supersymmetry transformation rules) receives corrections involving torsionful spin connections\footnote{These are the so-called \textit{anomaly-related} corrections  \cite{SSrestore} needed in order to restore supersymmetry broken after inclusion of Yang-Mills and Lorentz Chern-Simons terms in the effective action via the Green-Schwarz anomaly cancellation mechanism \cite{GS}.} 
$\Omega_{\pm} = \omega \pm H$ and their curvatures $\mathcal{R}_\pm$, where the axion curvature $H$ acts as a torsion to the Levi-Civita spin connection $\omega$.
The torsion then enters Ricci and Bianchi identities as well as other geometrical relations heavily involved in proving the auxiliary results needed for the general recurrent case. 

However, we will argue that at least certain types of $\alpha^\prime$ corrections still vanish. 
Firstly, corrections to $H$ at the order $\alpha^\prime$ are given by the Yang-Mills and Lorentz Chern-Simons forms $\omega_{YM}$ and $\omega_L$, respectively \cite{GS,guven}. One can easily verify that both $\omega_{YM}$ and $\omega_L$ stay zero for the extended class of solutions and consequently $H=\de B$ remains exact to all orders in $\alpha^\prime$. 
In turn, connections $\Omega_{\pm}$ receive no corrections either. Then, due to the nice behavior of $H$, the fields $\mathcal{R}, F,\nabla \phi, H$ and even $\mathcal{R}_{\pm}$ remain $VSI$ with respect to both types of derivatives $\nabla$ and $\nabla_\pm$ (associated with $\Omega_\pm$) 
in the sense that both types of derivatives are allowed in the construction of scalar polynomial invariants.  
We thus conclude that all corrections to \eqref{sugraaction} in the on-shell string effective action vanish. 
Similarly, any form at least quadratic in $\nabla \phi, F, H$ or involving their derivatives ($\nabla$ or $\nabla_\pm$) vanishes as well as any form constructed from $\mathcal{R}, \mathcal{R}_{\pm}$ and their derivatives of arbitrary order. 
Therefore, all $\alpha^\prime$ corrections of this nature to dilaton, axion and Yang-Mills equation necessarily vanish as well.

\subsection{Explicit form of the solutions}
In coordinates $(r,u,x^i)$, $i=2,\dots,10$, of section \ref{localform}, this class of solutions corresponds to 
 G{\"u}ven's solutions above with the plane wave metric replaced by a more general metric \eqref{VSImetric} satisfying \eqref{Weylcond} and \eqref{-1ricci}.  
This metric together with plane wave fields $A$, $\phi$, $B$ are then subject to a single Einstein equation
\begin{equation}\label{-2ricci2}
\begin{gathered}
 \Delta H^{(0)} =\  2 H\indices{^{(1)}_{,k}}W^{k} + H^{(1)}W\indices{^{k}_{,k}}  + W_{[m, n]}^2  \\
+ {W_{m,u}} \indices{^{m}} -2 (\phi^{\prime}/\phi)^2 - \kappa_\phi^2 f_{i}^a f^{a i} + 36\kappa_\phi^4 b_{[i,j]} b^{[i,j]}
\end{gathered}
\end{equation}
with $\kappa_\phi^2 \equiv \kappa^2 \phi^{-1}$. The rest of the field equations of the action \eqref{sugraaction} is automatically satisfied.
We thus end up with plane waves $A,\phi,B$ coupled with a broader class of spacetime metrics with nonvanishing $H^{(1)}, W_i $ and with $H^{(0)}$ whose dependence on the transverse coordinates is restricted only by the Einstein equation \eqref{-2ricci2}.

\begin{acknowledgments}
I would like to thank to Alena Pravdov{\'a} and Marcello Ortaggio for useful comments on the manuscript. 
This work has been supported by Research Plan RVO: 67985840 and Research Grant GAČR 19-09659S.
\end{acknowledgments}




\begin{thebibliography}{30}%
\makeatletter
\providecommand \@ifxundefined [1]{%
 \@ifx{#1\undefined}
}%
\providecommand \@ifnum [1]{%
 \ifnum #1\expandafter \@firstoftwo
 \else \expandafter \@secondoftwo
 \fi
}%
\providecommand \@ifx [1]{%
 \ifx #1\expandafter \@firstoftwo
 \else \expandafter \@secondoftwo
 \fi
}%
\providecommand \natexlab [1]{#1}%
\providecommand \enquote  [1]{``#1''}%
\providecommand \bibnamefont  [1]{#1}%
\providecommand \bibfnamefont [1]{#1}%
\providecommand \citenamefont [1]{#1}%
\providecommand \href@noop [0]{\@secondoftwo}%
\providecommand \href [0]{\begingroup \@sanitize@url \@href}%
\providecommand \@href[1]{\@@startlink{#1}\@@href}%
\providecommand \@@href[1]{\endgroup#1\@@endlink}%
\providecommand \@sanitize@url [0]{\catcode `\\12\catcode `\$12\catcode
  `\&12\catcode `\#12\catcode `\^12\catcode `\_12\catcode `\%12\relax}%
\providecommand \@@startlink[1]{}%
\providecommand \@@endlink[0]{}%
\providecommand \url  [0]{\begingroup\@sanitize@url \@url }%
\providecommand \@url [1]{\endgroup\@href {#1}{\urlprefix }}%
\providecommand \urlprefix  [0]{URL }%
\providecommand \Eprint [0]{\href }%
\providecommand \doibase [0]{https://doi.org/}%
\providecommand \selectlanguage [0]{\@gobble}%
\providecommand \bibinfo  [0]{\@secondoftwo}%
\providecommand \bibfield  [0]{\@secondoftwo}%
\providecommand \translation [1]{[#1]}%
\providecommand \BibitemOpen [0]{}%
\providecommand \bibitemStop [0]{}%
\providecommand \bibitemNoStop [0]{.\EOS\space}%
\providecommand \EOS [0]{\spacefactor3000\relax}%
\providecommand \BibitemShut  [1]{\csname bibitem#1\endcsname}%
\let\auto@bib@innerbib\@empty
\bibitem [{\citenamefont {G{\"u}ven}(1987)}]{guven}%
  \BibitemOpen
  \bibfield  {author} {\bibinfo {author} {\bibfnamefont {R.}~\bibnamefont
  {G{\"u}ven}},\ }\bibfield  {title} {\bibinfo {title} {Plane waves in
  effective field theories of superstrings},\ }\href@noop {} {\bibfield
  {journal} {\bibinfo  {journal} {Physics Letters B}\ }\textbf {\bibinfo
  {volume} {191}},\ \bibinfo {pages} {275} (\bibinfo {year}
  {1987})}\BibitemShut {NoStop}%
\bibitem [{\citenamefont {Amati}\ and\ \citenamefont
  {Klim{\v{c}}{\'\i}k}(1989)}]{klimcik}%
  \BibitemOpen
  \bibfield  {author} {\bibinfo {author} {\bibfnamefont {D.}~\bibnamefont
  {Amati}}\ and\ \bibinfo {author} {\bibfnamefont {C.}~\bibnamefont
  {Klim{\v{c}}{\'\i}k}},\ }\bibfield  {title} {\bibinfo {title}
  {Nonperturbative computation of the {Weyl} anomaly for a class of nontrivial
  backgrounds},\ }\href@noop {} {\bibfield  {journal} {\bibinfo  {journal}
  {Physics Letters B}\ }\textbf {\bibinfo {volume} {219}},\ \bibinfo {pages}
  {443} (\bibinfo {year} {1989})}\BibitemShut {NoStop}%
\bibitem [{\citenamefont {Horowitz}\ and\ \citenamefont
  {Steif}(1990)}]{horowitz}%
  \BibitemOpen
  \bibfield  {author} {\bibinfo {author} {\bibfnamefont {G.~T.}\ \bibnamefont
  {Horowitz}}\ and\ \bibinfo {author} {\bibfnamefont {A.~R.}\ \bibnamefont
  {Steif}},\ }\bibfield  {title} {\bibinfo {title} {Spacetime singularities in
  string theory},\ }\href@noop {} {\bibfield  {journal} {\bibinfo  {journal}
  {Physical Review Letters}\ }\textbf {\bibinfo {volume} {64}},\ \bibinfo
  {pages} {260} (\bibinfo {year} {1990})}\BibitemShut {NoStop}%
\bibitem [{\citenamefont {Coley}\ \emph {et~al.}(2008)\citenamefont {Coley},
  \citenamefont {Gibbons}, \citenamefont {Hervik},\ and\ \citenamefont
  {Pope}}]{metricsvanishing}%
  \BibitemOpen
  \bibfield  {author} {\bibinfo {author} {\bibfnamefont {A.}~\bibnamefont
  {Coley}}, \bibinfo {author} {\bibfnamefont {G.}~\bibnamefont {Gibbons}},
  \bibinfo {author} {\bibfnamefont {S.}~\bibnamefont {Hervik}},\ and\ \bibinfo
  {author} {\bibfnamefont {C.}~\bibnamefont {Pope}},\ }\bibfield  {title}
  {\bibinfo {title} {Metrics with vanishing quantum corrections},\ }\href@noop
  {} {\bibfield  {journal} {\bibinfo  {journal} {Classical and Quantum
  Gravity}\ }\textbf {\bibinfo {volume} {25}},\ \bibinfo {pages} {145017}
  (\bibinfo {year} {2008})}\BibitemShut {NoStop}%
\bibitem [{\citenamefont {Hervik}\ \emph {et~al.}(2014)\citenamefont {Hervik},
  \citenamefont {Pravda},\ and\ \citenamefont
  {Pravdov{\'a}}}]{typeIIINuniversal}%
  \BibitemOpen
  \bibfield  {author} {\bibinfo {author} {\bibfnamefont {S.}~\bibnamefont
  {Hervik}}, \bibinfo {author} {\bibfnamefont {V.}~\bibnamefont {Pravda}},\
  and\ \bibinfo {author} {\bibfnamefont {A.}~\bibnamefont {Pravdov{\'a}}},\
  }\bibfield  {title} {\bibinfo {title} {Type {III} and {N} universal
  spacetimes},\ }\href@noop {} {\bibfield  {journal} {\bibinfo  {journal}
  {Classical and Quantum Gravity}\ }\textbf {\bibinfo {volume} {31}},\ \bibinfo
  {pages} {215005} (\bibinfo {year} {2014})}\BibitemShut {NoStop}%
\bibitem [{\citenamefont {Ortaggio}\ and\ \citenamefont
  {Pravda}(2016)}]{VSIelmag}%
  \BibitemOpen
  \bibfield  {author} {\bibinfo {author} {\bibfnamefont {M.}~\bibnamefont
  {Ortaggio}}\ and\ \bibinfo {author} {\bibfnamefont {V.}~\bibnamefont
  {Pravda}},\ }\bibfield  {title} {\bibinfo {title} {Electromagnetic fields
  with vanishing scalar invariants},\ }\href@noop {} {\bibfield  {journal}
  {\bibinfo  {journal} {Classical and Quantum Gravity}\ }\textbf {\bibinfo
  {volume} {33}},\ \bibinfo {pages} {115010} (\bibinfo {year}
  {2016})}\BibitemShut {NoStop}%
\bibitem [{\citenamefont {Ortaggio}\ and\ \citenamefont
  {Pravda}(2018)}]{quantumelmag}%
  \BibitemOpen
  \bibfield  {author} {\bibinfo {author} {\bibfnamefont {M.}~\bibnamefont
  {Ortaggio}}\ and\ \bibinfo {author} {\bibfnamefont {V.}~\bibnamefont
  {Pravda}},\ }\bibfield  {title} {\bibinfo {title} {Electromagnetic fields
  with vanishing quantum corrections},\ }\href@noop {} {\bibfield  {journal}
  {\bibinfo  {journal} {Physics Letters B}\ }\textbf {\bibinfo {volume}
  {779}},\ \bibinfo {pages} {393} (\bibinfo {year} {2018})}\BibitemShut
  {NoStop}%
\bibitem [{\citenamefont {Hervik}\ \emph {et~al.}(2018)\citenamefont {Hervik},
  \citenamefont {Ortaggio},\ and\ \citenamefont {Pravda}}]{universalMaxwell}%
  \BibitemOpen
  \bibfield  {author} {\bibinfo {author} {\bibfnamefont {S.}~\bibnamefont
  {Hervik}}, \bibinfo {author} {\bibfnamefont {M.}~\bibnamefont {Ortaggio}},\
  and\ \bibinfo {author} {\bibfnamefont {V.}~\bibnamefont {Pravda}},\
  }\bibfield  {title} {\bibinfo {title} {Universal electromagnetic fields},\
  }\href@noop {} {\bibfield  {journal} {\bibinfo  {journal} {Classical and
  Quantum Gravity}\ }\textbf {\bibinfo {volume} {35}},\ \bibinfo {pages}
  {175017} (\bibinfo {year} {2018})}\BibitemShut {NoStop}%
\bibitem [{\citenamefont {Kuchynka}\ and\ \citenamefont
  {Ortaggio}(2019)}]{EMuniversal}%
  \BibitemOpen
  \bibfield  {author} {\bibinfo {author} {\bibfnamefont {M.}~\bibnamefont
  {Kuchynka}}\ and\ \bibinfo {author} {\bibfnamefont {M.}~\bibnamefont
  {Ortaggio}},\ }\bibfield  {title} {\bibinfo {title} {{Einstein-Maxwell}
  solutions with vanishing higher-order corrections},\ }\href@noop {}
  {\bibfield  {journal} {\bibinfo  {journal} {Physical Review D}\ }\textbf
  {\bibinfo {volume} {99}} (\bibinfo {year} {2019})}\BibitemShut {NoStop}%
\bibitem [{\citenamefont {Kuchynka}(2019)}]{YMuniversal}%
  \BibitemOpen
  \bibfield  {author} {\bibinfo {author} {\bibfnamefont {M.}~\bibnamefont
  {Kuchynka}},\ }\bibfield  {title} {\bibinfo {title} {Gauge fields with
  vanishing scalar invariants},\ }\href@noop {} {\bibfield  {journal} {\bibinfo
   {journal} {Classical and Quantum Gravity}\ }\textbf {\bibinfo {volume}
  {36}},\ \bibinfo {pages} {205016} (\bibinfo {year} {2019})}\BibitemShut
  {NoStop}%
\bibitem [{\citenamefont {Tseytlin}(1997)}]{BITseytlin}%
  \BibitemOpen
  \bibfield  {author} {\bibinfo {author} {\bibfnamefont {A.~A.}\ \bibnamefont
  {Tseytlin}},\ }\bibfield  {title} {\bibinfo {title} {On non-abelian
  generalisation of {Born}-{Infeld} action in string theory},\ }\href@noop {}
  {\bibfield  {journal} {\bibinfo  {journal} {Nucl. Phys. B}\ }\textbf
  {\bibinfo {volume} {501}},\ \bibinfo {pages} {41} (\bibinfo {year}
  {1997})}\BibitemShut {NoStop}%
\bibitem [{\citenamefont {Tchrakian}(1993)}]{YMlovelock2}%
  \BibitemOpen
  \bibfield  {author} {\bibinfo {author} {\bibfnamefont {D.}~\bibnamefont
  {Tchrakian}},\ }\bibfield  {title} {\bibinfo {title} {Yang-{M}ills
  hierarchy},\ }\href@noop {} {\bibfield  {journal} {\bibinfo  {journal}
  {Differential Geometric Methods in Theoretical Physics (Singapore: World
  Scientific), (https://doi.org/10.1142/9789814536448)}\ } (\bibinfo {year}
  {1993})}\BibitemShut {NoStop}%
\bibitem [{\citenamefont {Muller-Hoissen}(1988)}]{nonminYM2}%
  \BibitemOpen
  \bibfield  {author} {\bibinfo {author} {\bibfnamefont {F.}~\bibnamefont
  {Muller-Hoissen}},\ }\bibfield  {title} {\bibinfo {title} {Modification of
  {Einstein}-{Yang}-{Mills} theory from dimensional reduction of the
  gauss-bonnet action},\ }\href@noop {} {\bibfield  {journal} {\bibinfo
  {journal} {Classical and Quantum Gravity}\ }\textbf {\bibinfo {volume} {5}},\
  \bibinfo {pages} {L35} (\bibinfo {year} {1988})}\BibitemShut {NoStop}%
\bibitem [{\citenamefont {Balakin}\ and\ \citenamefont
  {Zayats}(2007)}]{nonminYM1}%
  \BibitemOpen
  \bibfield  {author} {\bibinfo {author} {\bibfnamefont {A.~B.}\ \bibnamefont
  {Balakin}}\ and\ \bibinfo {author} {\bibfnamefont {A.~E.}\ \bibnamefont
  {Zayats}},\ }\bibfield  {title} {\bibinfo {title} {Non-minimal {Wu}--{Yang}
  monopole},\ }\href@noop {} {\bibfield  {journal} {\bibinfo  {journal}
  {Physics Letters B}\ }\textbf {\bibinfo {volume} {644}},\ \bibinfo {pages}
  {294} (\bibinfo {year} {2007})}\BibitemShut {NoStop}%
\bibitem [{\citenamefont {Coley}\ \emph {et~al.}(2004)\citenamefont {Coley},
  \citenamefont {Milson}, \citenamefont {Pravda},\ and\ \citenamefont
  {Pravdov{\'a}}}]{VSIinHD}%
  \BibitemOpen
  \bibfield  {author} {\bibinfo {author} {\bibfnamefont {A.}~\bibnamefont
  {Coley}}, \bibinfo {author} {\bibfnamefont {R.}~\bibnamefont {Milson}},
  \bibinfo {author} {\bibfnamefont {V.}~\bibnamefont {Pravda}},\ and\ \bibinfo
  {author} {\bibfnamefont {A.}~\bibnamefont {Pravdov{\'a}}},\ }\bibfield
  {title} {\bibinfo {title} {Vanishing scalar invariant spacetimes in higher
  dimensions},\ }\href@noop {} {\bibfield  {journal} {\bibinfo  {journal}
  {Classical and Quantum Gravity}\ }\textbf {\bibinfo {volume} {21}},\ \bibinfo
  {pages} {5519} (\bibinfo {year} {2004})}\BibitemShut {NoStop}%
\bibitem [{\citenamefont {Coley}\ \emph {et~al.}(2006)\citenamefont {Coley},
  \citenamefont {Fuster}, \citenamefont {Hervik},\ and\ \citenamefont
  {Pelavas}}]{HDVSI}%
  \BibitemOpen
  \bibfield  {author} {\bibinfo {author} {\bibfnamefont {A.}~\bibnamefont
  {Coley}}, \bibinfo {author} {\bibfnamefont {A.}~\bibnamefont {Fuster}},
  \bibinfo {author} {\bibfnamefont {S.}~\bibnamefont {Hervik}},\ and\ \bibinfo
  {author} {\bibfnamefont {N.}~\bibnamefont {Pelavas}},\ }\bibfield  {title}
  {\bibinfo {title} {Higher dimensional {VSI} spacetimes},\ }\href@noop {}
  {\bibfield  {journal} {\bibinfo  {journal} {Classical and Quantum Gravity}\
  }\textbf {\bibinfo {volume} {23}},\ \bibinfo {pages} {7431} (\bibinfo {year}
  {2006})}\BibitemShut {NoStop}%
\bibitem [{\citenamefont {Ortaggio}\ \emph {et~al.}(2013)\citenamefont
  {Ortaggio}, \citenamefont {Pravda},\ and\ \citenamefont
  {Pravdov{\'a}}}]{review}%
  \BibitemOpen
  \bibfield  {author} {\bibinfo {author} {\bibfnamefont {M.}~\bibnamefont
  {Ortaggio}}, \bibinfo {author} {\bibfnamefont {V.}~\bibnamefont {Pravda}},\
  and\ \bibinfo {author} {\bibfnamefont {A.}~\bibnamefont {Pravdov{\'a}}},\
  }\bibfield  {title} {\bibinfo {title} {Algebraic classification of higher
  dimensional spacetimes based on null alignment},\ }\href@noop {} {\bibfield
  {journal} {\bibinfo  {journal} {Classical and Quantum Gravity}\ }\textbf
  {\bibinfo {volume} {30}},\ \bibinfo {pages} {013001} (\bibinfo {year}
  {2013})}\BibitemShut {NoStop}%
\bibitem [{\citenamefont {G{\"u}ven}(1979)}]{guvenEYM}%
  \BibitemOpen
  \bibfield  {author} {\bibinfo {author} {\bibfnamefont {R.}~\bibnamefont
  {G{\"u}ven}},\ }\bibfield  {title} {\bibinfo {title} {Solution for gravity
  coupled to non-abelian plane waves},\ }\href@noop {} {\bibfield  {journal}
  {\bibinfo  {journal} {Physical Review D}\ }\textbf {\bibinfo {volume} {19}},\
  \bibinfo {pages} {471} (\bibinfo {year} {1979})}\BibitemShut {NoStop}%
\bibitem [{\citenamefont {Chapline}\ and\ \citenamefont
  {Manton}(1989)}]{chapline}%
  \BibitemOpen
  \bibfield  {author} {\bibinfo {author} {\bibfnamefont {G.~F.}\ \bibnamefont
  {Chapline}}\ and\ \bibinfo {author} {\bibfnamefont {N.~S.}\ \bibnamefont
  {Manton}},\ }\bibfield  {title} {\bibinfo {title} {Unification of
  {Yang}-{Mills} theory and supergravity in ten dimensions},\ }in\ \href@noop
  {} {\emph {\bibinfo {booktitle} {Supergravities in Diverse Dimensions:
  Commentary and Reprints (In 2 Volumes)}}}\ (\bibinfo  {publisher} {World
  Scientific},\ \bibinfo {year} {1989})\ pp.\ \bibinfo {pages}
  {200--204}\BibitemShut {NoStop}%
\bibitem [{\citenamefont {Figueroa-O'Farrill}\ \emph
  {et~al.}(2007)\citenamefont {Figueroa-O'Farrill}, \citenamefont
  {Hackett-Jones},\ and\ \citenamefont {Moutsopoulos}}]{farrill}%
  \BibitemOpen
  \bibfield  {author} {\bibinfo {author} {\bibfnamefont {J.}~\bibnamefont
  {Figueroa-O'Farrill}}, \bibinfo {author} {\bibfnamefont {E.}~\bibnamefont
  {Hackett-Jones}},\ and\ \bibinfo {author} {\bibfnamefont {G.}~\bibnamefont
  {Moutsopoulos}},\ }\bibfield  {title} {\bibinfo {title} {The {Killing}
  superalgebra of 10-dimensional supergravity backgrounds},\ }\href@noop {}
  {\bibfield  {journal} {\bibinfo  {journal} {Classical and Quantum Gravity}\
  }\textbf {\bibinfo {volume} {24}},\ \bibinfo {pages} {3291} (\bibinfo {year}
  {2007})}\BibitemShut {NoStop}%
\bibitem [{\citenamefont {Coley}\ \emph {et~al.}(2007)\citenamefont {Coley},
  \citenamefont {Fuster}, \citenamefont {Hervik},\ and\ \citenamefont
  {Pelavas}}]{VSIsugra}%
  \BibitemOpen
  \bibfield  {author} {\bibinfo {author} {\bibfnamefont {A.}~\bibnamefont
  {Coley}}, \bibinfo {author} {\bibfnamefont {A.}~\bibnamefont {Fuster}},
  \bibinfo {author} {\bibfnamefont {S.}~\bibnamefont {Hervik}},\ and\ \bibinfo
  {author} {\bibfnamefont {N.}~\bibnamefont {Pelavas}},\ }\bibfield  {title}
  {\bibinfo {title} {Vanishing scalar invariant spacetimes in supergravity},\
  }\href@noop {} {\bibfield  {journal} {\bibinfo  {journal} {Journal of High
  Energy Physics}\ }\textbf {\bibinfo {volume} {2007}},\ \bibinfo {pages} {032}
  (\bibinfo {year} {2007})}\BibitemShut {NoStop}%
\bibitem [{\citenamefont {Fulling}\ \emph {et~al.}(1992)\citenamefont
  {Fulling}, \citenamefont {King}, \citenamefont {Wybourne},\ and\
  \citenamefont {Cummins}}]{fkwcpaper}%
  \BibitemOpen
  \bibfield  {author} {\bibinfo {author} {\bibfnamefont {S.}~\bibnamefont
  {Fulling}}, \bibinfo {author} {\bibfnamefont {R.~C.}\ \bibnamefont {King}},
  \bibinfo {author} {\bibfnamefont {B.}~\bibnamefont {Wybourne}},\ and\
  \bibinfo {author} {\bibfnamefont {C.}~\bibnamefont {Cummins}},\ }\bibfield
  {title} {\bibinfo {title} {Normal forms for tensor polynomials. {I.} {T}he
  {Riemann} tensor},\ }\href@noop {} {\bibfield  {journal} {\bibinfo  {journal}
  {Classical and Quantum Gravity}\ }\textbf {\bibinfo {volume} {9}},\ \bibinfo
  {pages} {1151} (\bibinfo {year} {1992})}\BibitemShut {NoStop}%
\bibitem [{\citenamefont {Coley}\ \emph {et~al.}(2009)\citenamefont {Coley},
  \citenamefont {Hervik}, \citenamefont {Papadopoulos},\ and\ \citenamefont
  {Pelavas}}]{hervikKundt}%
  \BibitemOpen
  \bibfield  {author} {\bibinfo {author} {\bibfnamefont {A.}~\bibnamefont
  {Coley}}, \bibinfo {author} {\bibfnamefont {S.}~\bibnamefont {Hervik}},
  \bibinfo {author} {\bibfnamefont {G.}~\bibnamefont {Papadopoulos}},\ and\
  \bibinfo {author} {\bibfnamefont {N.}~\bibnamefont {Pelavas}},\ }\bibfield
  {title} {\bibinfo {title} {Kundt spacetimes},\ }\href@noop {} {\bibfield
  {journal} {\bibinfo  {journal} {Classical and Quantum Gravity}\ }\textbf
  {\bibinfo {volume} {26}},\ \bibinfo {pages} {105016} (\bibinfo {year}
  {2009})}\BibitemShut {NoStop}%
\bibitem [{\citenamefont {Kuchynka}\ \emph {et~al.}(2019)\citenamefont
  {Kuchynka}, \citenamefont {M{\'a}lek}, \citenamefont {V},\ and\ \citenamefont
  {Pravdov{\'a}}}]{Gorilla}%
  \BibitemOpen
  \bibfield  {author} {\bibinfo {author} {\bibfnamefont {M.}~\bibnamefont
  {Kuchynka}}, \bibinfo {author} {\bibfnamefont {T.}~\bibnamefont {M{\'a}lek}},
  \bibinfo {author} {\bibfnamefont {P.}~\bibnamefont {V}},\ and\ \bibinfo
  {author} {\bibfnamefont {A.}~\bibnamefont {Pravdov{\'a}}},\ }\bibfield  {title}
  {\bibinfo {title} {Almost universal spacetimes in higher-order gravity
  theories},\ }\href@noop {} {\bibfield  {journal} {\bibinfo  {journal}
  {Physical Review D}\ }\textbf {\bibinfo {volume} {99}},\ \bibinfo {pages}
  {024043} (\bibinfo {year} {2019})}\BibitemShut {NoStop}%
\bibitem [{\citenamefont {G{\"u}rses}\ \emph {et~al.}(2013)\citenamefont
  {G{\"u}rses}, \citenamefont {Hervik}, \citenamefont {{\c{S}}i{\c{s}}man},\
  and\ \citenamefont {Tekin}}]{gurses2013}%
  \BibitemOpen
  \bibfield  {author} {\bibinfo {author} {\bibfnamefont {M.}~\bibnamefont
  {G{\"u}rses}}, \bibinfo {author} {\bibfnamefont {S.}~\bibnamefont {Hervik}},
  \bibinfo {author} {\bibfnamefont {T.}~\bibnamefont {{\c{S}}i{\c{s}}man}},\
  and\ \bibinfo {author} {\bibfnamefont {B.}~\bibnamefont {Tekin}},\ }\bibfield
   {title} {\bibinfo {title} {Anti--de {Sitter}--{Wave} solutions of higher
  derivative theories},\ }\href@noop {} {\bibfield  {journal} {\bibinfo
  {journal} {Physical review letters}\ }\textbf {\bibinfo {volume} {111}},\
  \bibinfo {pages} {101101} (\bibinfo {year} {2013})}\BibitemShut {NoStop}%
\bibitem [{\citenamefont {Fuster}\ and\ \citenamefont {van
  Holten}(2005)}]{typeIIIEYM}%
  \BibitemOpen
  \bibfield  {author} {\bibinfo {author} {\bibfnamefont {A.}~\bibnamefont
  {Fuster}}\ and\ \bibinfo {author} {\bibfnamefont {J.-W.}\ \bibnamefont {van
  Holten}},\ }\bibfield  {title} {\bibinfo {title} {Type {III}
  {Einstein}-{Yang}-{Mills} solutions},\ }\href@noop {} {\bibfield  {journal}
  {\bibinfo  {journal} {Physical Review D}\ }\textbf {\bibinfo {volume} {72}},\
  \bibinfo {pages} {024011} (\bibinfo {year} {2005})}\BibitemShut {NoStop}%
\bibitem [{\citenamefont {Horowitz}\ and\ \citenamefont
  {Tseytlin}(1995)}]{hortsey}%
  \BibitemOpen
  \bibfield  {author} {\bibinfo {author} {\bibfnamefont {G.~T.}\ \bibnamefont
  {Horowitz}}\ and\ \bibinfo {author} {\bibfnamefont {A.~A.}\ \bibnamefont
  {Tseytlin}},\ }\bibfield  {title} {\bibinfo {title} {New class of exact
  solutions in string theory},\ }\href@noop {} {\bibfield  {journal} {\bibinfo
  {journal} {Physical Review D}\ }\textbf {\bibinfo {volume} {51}},\ \bibinfo
  {pages} {2896} (\bibinfo {year} {1995})}\BibitemShut {NoStop}%
\bibitem [{\citenamefont {Tseytlin}(1995)}]{TseytlinDuality}%
  \BibitemOpen
  \bibfield  {author} {\bibinfo {author} {\bibfnamefont {A.~A.}\ \bibnamefont
  {Tseytlin}},\ }\bibfield  {title} {\bibinfo {title} {Exact string solutions
  and duality},\ }in\ \href@noop {} {\emph {\bibinfo {booktitle} {Second Paris
  Cosmology Colloquium}}}\ (\bibinfo {year} {1995})\ p.\ \bibinfo {pages}
  {371}\BibitemShut {NoStop}%
\bibitem [{\citenamefont {Bergshoeff}\ and\ \citenamefont
  {De~Roo}(1989)}]{SSrestore}%
  \BibitemOpen
  \bibfield  {author} {\bibinfo {author} {\bibfnamefont {E.}~\bibnamefont
  {Bergshoeff}}\ and\ \bibinfo {author} {\bibfnamefont {M.}~\bibnamefont
  {De~Roo}},\ }\bibfield  {title} {\bibinfo {title} {The quartic effective
  action of the heterotic string and supersymmetry},\ }\href@noop {} {\bibfield
   {journal} {\bibinfo  {journal} {Nuclear Physics B}\ }\textbf {\bibinfo
  {volume} {328}},\ \bibinfo {pages} {439} (\bibinfo {year}
  {1989})}\BibitemShut {NoStop}%
\bibitem [{\citenamefont {Green}\ and\ \citenamefont {Schwarz}(1984)}]{GS}%
  \BibitemOpen
  \bibfield  {author} {\bibinfo {author} {\bibfnamefont {M.~B.}\ \bibnamefont
  {Green}}\ and\ \bibinfo {author} {\bibfnamefont {J.~H.}\ \bibnamefont
  {Schwarz}},\ }\bibfield  {title} {\bibinfo {title} {Anomaly cancellations in
  supersymmetric d= 10 gauge theory and superstring theory},\ }\href@noop {}
  {\bibfield  {journal} {\bibinfo  {journal} {Physics Letters B}\ }\textbf
  {\bibinfo {volume} {149}},\ \bibinfo {pages} {117} (\bibinfo {year}
  {1984})}\BibitemShut {NoStop}%
\end{thebibliography}

%

\end{document}